\documentstyle[12pt]{article}
\def\sfrac#1#2{{\scriptstyle#1\over\scriptstyle#2}}
\def\frac#1#2{{\displaystyle#1\over\displaystyle#2}}

\begin{document}
\begin{center}
{\bf MACHO HALO OBJECTS AS  SOURCES OF GAMMA BURSTS}
\footnote{Accepted for publication in Phys.Lett.A (1997)}
\vspace{5mm}
A.V.Gurevich and K.P.Zybin\\
\vspace{5mm}
{\em  P.N.Lebedev Institute of Physics, Russian Academy of Sciences,
Moscow, Russia}\\
\end{center}
\begin{abstract}
A new model of gamma bursts (GB)  generated by MACHO's in the form of
noncompact dark matter halo objects (NO) is proposed. The model explain well
the spherical symmetry, $log N - log S$ curve of GB absolutely identically
to the well-known giant dark matter halo (GDMH) model. The main difficulty of
GDMH model -- the lack of neutron stars in our galactic halo is eliminated
because of the large number of
NO ($\sim 10^{13}$) which is quite close to a total
number of bursts during the life time of the Universe. The proposed mechanism
of the burst is the instability and explosion of the baryonic core (BC) of NO.
The structure of BC and possible mechanism of BC explosion are discussed.

The intensive infrared emission from BC of neutralino stars (NeS) at the
wavelength $(3 - 8)\mu$ is predicted. The possibility to discover NeS by
observation of infrared emission from the recently distinguished point-like
gamma sources is indicated.
\end{abstract}
\section{Introduction}
The origin of gamma bursts is one of the mostly intriguing problem
in astrophysics during the last 20 years. Analysis of the
observational data shows that today exist two main models of the
origin of gamma bursts: cosmological and giant halo model [1].

Giant dark matter halo (GDMH) model was proposed in [2]. It
considers relic neutron stars as the source of gamma bursts.
These stars should have the same density distribution as
distribution of cold dark matter in the Galactic halo, which has a
size $R_h\approx 200$kpc [2,3]. This proposition
allowed to explain the main statistical properties of gamma
bursts: their tightly spherical symmetry and a considerable condensation
to the center, known as $log N - log S$ curve [1].  Furthermore
GDMH model fairly well describes weak asymmetries in the distribution
of gamma-ray bursts, observed during first years of measurement
 by the BATSE device of COMPTON Observatory [4,5]. The characteristic
energy $E_{\gamma}$ released in  gamma-rays in one burst as follows
from this model is $10^{41}-10^{42}$ erg.

The  weak point of this model is the supposition that the source
of gamma bursts is relic neutron stars (NS). The full number of these
stars  is not large enough $N\leq 10^6-10^7$, what means that
every NS must repeat gamma burst many times: $10^6-10^7$ , since
the total number of bursts should be $\sim 10^{13}$. The same
difficulty has analogous model, which consider high velocity
NS as a source of   gamma bursts [6,7]. These stars have got high
velocities during supernova explosion and are moving out of the
Galaxy. The number of such NS is about $10^6 - 10^7$. It is difficult to
find  the mechanism of repeating gamma bursts from NS and the source of
energy needed for repeating $10^6$ times bursts. It should be
noted also that according to observations bursts never repeat (we do not
speak here about special small group of the so-called "repeaters").
These difficulties for the NS models as the source of gamma bursts
were emphasized in numerous papers [2-9].

Here we will propose a new mechanism of gamma burst connected with the
instability and a possible  explosion of the baryonic core of
noncompact dark matter halo objects (NO).
Microlensing  established, that the large part of
dark matter in Galaxy halo consist of MACHO's --
unseen objects with masses
$(0.05-0.8)M_{\odot}$ [10,11]. These objects are naturally considered
to be White Dwarfs, Brown Dwarfs, Jupiters or some other stars
or planets. To the contrary in [12,13] was supposed the existence of
quite a new type of massive structures -- gravitationally
compressed objects from nonbaryonic cold dark matter (CDM).
These objects are the results of a small scale hierarchical structuring
developing in CDM [14,15]. They are noncompact, having some spectrum of
characteristic dimensions and masses.
After recombination a
baryonic component falls down into the potential well formed by
nonbaryonic matter. In this way the compact baryonic core
is created. The nonbaryonic component forms a
spread spherical halo, where the dominant mass of the object
is concentrated. The main hypothesis of [12]-[15] is that these
noncompact nonbaryonic objects (NO) determine significant part of
microlensing events observed in [11].

NO were proposed in [12,13] to avoid the difficulties arising
from microlensing observations. Here we consider them as the source of
gamma bursts. Because of supposition that the main
part of dark matter consist of NO
this mechanism gives the same statistical
results as GDMH model. It explains well the spherical symmetry,
$log N - log S$ curve and small asymmetry absolutely identically to [2-5].
The new very essential point is connected with $N_{NO}$ -- the number of NO
in Galaxy halo. According to [12-15] $N_{NO}\sim 10^{13}$ what is quite close to
the total number of bursts in our Galaxy during lifetime of the
Universe. So no bursts repetition is needed for this model.

\section{Dark matter in the baryonic core.}
Non dissipative cold dark matter creates stationary spherically-symmetrical
structures compressed by their own gravitational forces. For these objects
the theory established a fundamental scaling low for density distribution [16]:
\begin{equation}
\rho\propto r^{-\alpha},\qquad \qquad \alpha\approx 1.8
\end{equation}
which holds quite well for large scale structures in the Universe.

Noncompact dark matter objects (NO) are the small scale structures with the
masses $M_x\sim (0.01-1)M_{\odot}$ and characteristic dimensions $R_x\sim
10^{14} - 10^{15}$ cm [13-15]. For small scale dark matter structures the
distribution (1) holds on up to the dimensions $r_c$  an order magnitude less
than $R_x$ [17]:
\begin{equation}
r_c\sim 0.1 R_x.
\end{equation}
The density
distribution of CDM particles in NO could be presented in a form [17]:
\begin{equation}
\rho=\left\{
\begin{array}{cc}
\rho_0 &\qquad \qquad 0<r<r_c \\
\rho_0 \left(\frac{r}{r_c}\right)^{-\alpha} &\qquad \qquad r_c<r<R_x \\
 0 &\qquad \qquad  r>R_x \\
\end{array}
\right.
\end{equation}

This distribution does not take into account the existence of baryonic core.
The fundamental difference between nonbaryonic dark matter and baryonic is
determined by dissipative processes. After recombination the baryonic matter
under the action of radiative cooling loses its thermal energy and falls
down into the potential wells formed by nonbaryonic matter. It leads to the
creation of baryonic body (baryonic core) with mass $M_b\sim 0.05 M_x$ at the
center of the dark matter object.

It is essential, that the process of the baryonic matter condensation change
significantly the density and a full amount of the nonbaryonic matter trapped
inside the baryonic core (BC). Let us consider this process in details. One
can neglect the collisions between nonbaryonic particles, which are oscillating
in the potential well created by their gravitational field.
It is easy to estimate, that the typical oscillation time ($\sim 10$ years)
is essentially less than the time of baryonic core formation.
It means that the core formation is a slow process.
 Because of this reason we can
describe the changing of distribution function of dark matter
particles using the adiabatic approximation.
Initial adiabatic invariant $I_i$ is defined by the formulae:
\begin{equation}
I_i=\int_{r_{min}}^{r_{max}} \sqrt{\hat{E}-
p_0 r^2 - \frac{m^2}{2r^2}}
\, dr \qquad \qquad \hat{E} = E - \psi_{00} \qquad p_0=\frac{2}{3}\pi G \rho_0
\end{equation}
Here $E=\varepsilon/m_x$, $\varepsilon$-is the
energy of a particle moving in potential
$p_0 r^2$, $\psi_{00}$- is the depth of the potential well,
$m$- is the particle angular momentum and $r_{min}, r_{max}$
are reflection points, determined as a roots of the expression under the
integral. Since we are interested in changing of distribution function
near BC, in (4) was chosen (in accordance with (3)) $\psi=\psi_{00}+
2\pi G\rho_0/3 r^2$.

After the formation of BC the adiabatic invariant
$I_f$ of a particle near the center takes a form
\begin{equation}
I_f=\int_{r_{min}}^{r_{max}} \sqrt{E+\frac{GM_b}{r} - \frac{m^2}{2r^2}}
\, dr
\end{equation}
here $GM_b/r$ -is a potential of a created baryonic body. In expression (5) we
took into account only potential of baryonic body and neglected the potential
of dark matter. As we will see below (12)
the mass of CDM particles in  BC and its
neighborhood is essentially less than the mass of baryonic body.

In view of conservation of the adiabatic invariant, it
follows that for any value of $I$, we have for the distribution function
$f(I)$:
$$
f(I)=f_0(I_i)|_{I_i=I}.
$$
The density  distribution $\rho$ in variables $E, m^2$ can be
written in a form [16]:
\begin{equation}
\rho(r)=\frac{2\pi m_x}{r^2}\int_{0}^{\infty} dm^2
\int_{\psi+\sfrac{m^2}{2r^2}}^{0}\frac{f(E)}{\sqrt{E-\psi-
\frac{m^2}{2r^2}}} dE
\end{equation}
It is natural to choose the initial distribution function $f_0$
as a maxwellian one
\begin{equation}
f_0=\frac{n_0}{(2\pi T)^{3/2}} e^{-E/T} , \qquad \qquad
T\approx \frac{G M_x}{R_x} \qquad\qquad n_0=\frac{\rho_0}{m_x}
\end{equation}
After integration of (6),(7) , one can find
\begin{equation}
I_i=\frac{\pi}{2^{3/2}}\left(\sqrt{\frac{2}{p_0}}\hat{E}- m\right)
, \qquad\qquad
I_f=\frac{\pi}{2}\left(\frac{2GM_b}{(-E)^{1/2}}-2^{1/2} m \right)
\end{equation}
Substituting (8) into (6) we obtain:
\begin{equation}
\rho(r)=\frac{2\pi m_x}{r^2}\int_{0}^{\infty} dm^2
\int_{-\sfrac{GM_b}{r}+\sfrac{m^2}{2r^2}}^{0}
\frac{n_0 e^{-\sfrac{\psi_{00}}{T}}}{(2\pi T)^{3/2}}\times
\frac{exp\left\{-\frac{2p_0^{1/2} GM_b}{T(-E)^{1/2}}+\frac{m p_0^{1/2}}
{2^{1/2}T}\right\}}{\sqrt{E-\psi-
\frac{m^2}{2r^2}}} dE
\end{equation}
Let us find the asymptotic of solution (9) in the limit $r\to\ 0$.
In this case $E\to -\infty$, and taking into account, that
$m^2 \leq 2GM_b r $, we find:
\begin{equation}
\rho(r)=\frac{8}{3\sqrt{2\pi}} n_0 e^{-\sfrac{\psi_{00}}{T}}
m_x \left(\frac{GM_b}{Tr}\right)^{3/2}
\end{equation}
It is easy to see that the law for density distribution (10)
does not depend on the shape of
distribution function (7), but it depends on its
 behavior at $E\to 0$
The law (10) has a singularity at $r\to 0$.  The behavior
of density  is defined by this singularity of potential,
but in reality the singular law (10) is cut on the scale
of baryonic core $r_b$.  Borrowing in mind (in accordance with (3)) that
\begin{equation}
\frac{4}{3}\pi n_0 m_x r_c^3 \approx 0.1 M_x,
\end{equation}
and choosing $T\approx\psi_{00}$ one can find from (10),(11):
$$
\rho(r_b)=33\times n_0 m_x \left(\frac{M_b r_c}{M_x r_b}\right)^{3/2}
$$

So we see, that the adiabatic capture of dark matter particles lead to
the increase of the
dark matter density  inside the baryonic core in $ 10^3 - 10^4$
times. Because of this  the full amount of the mass of a dark matter
trapped inside the BC $M_{xb}$ could reach a significant part:
\begin{equation}
\frac{M_{xb}}{M_x} \sim 10^{-6} - 10^{-7}
\end{equation}
It means for $M_x=0.5M_{\odot}$, that $M_{xb}\sim 10^{25} - 10^{26}$ g.

\section{The structure of baryonic core with CDM component.}

 Let us consider now the structure of baryonic core.
The stationary state of this core is described by equations:
\begin{eqnarray}
& \nabla P=-\nabla\psi \nonumber \\
& \nabla\left(\kappa\nabla T\right)= - \sigma T_s^4 S + Q_n + Q_{th} \\
& \Delta\psi=4\pi G \left(\rho_b +\rho_n\right) \nonumber
\end{eqnarray}
Here $P=P_b+P_n$ is the total pressure of baryons and nonbaryons,
$T(r)$ is the temperature of baryonic particles, $\psi$- potential of
gravitational field, $\kappa$-heat conductivity, $\rho_b$ and $\rho_n$
are densities of baryons and nonbaryons.
$Q_n$ and $Q_{th}$ are
the sources of the heating of BC by DM particles. These sources are
significant for the DM particles of a special type and would be considered
in the next section. $T_s$ is the temperature of a core surface $S$, $\sigma$
- Stephan-Boltzman constant.

The distribution of CDM particles inside the baryonic core is determined by
their collisions with baryons,it means that in stationary conditions they have
Boltzman distribution:
\begin{equation}
\rho_n=\rho_{n0}\,
exp \left\{-\frac{\psi(r) m_x}{T^{\star}}\right\}
\end{equation}
where $\psi(r)$ is gravitational potential and
$T^{\star}$ is the effective  temperature of CDM particles.
\begin{equation}
T^{\star}= \frac{3 m_x}{R_x^3}\int_{0}^{R_x} r^2 \left\{\frac{
\int_{\psi+\sfrac{m^2}{2r^2}}^0\frac{E f(E)
dE}{\sqrt{E-\psi-m^2/2r^2}}}{
\int_{\psi+\sfrac{m^2}{2r^2}}^0\frac{f(E) dE}{\sqrt{E-\psi-m^2/2r^2}}}
\right\}\,dr
\end{equation}
Here $\rho_{n0}$ is  CDM density in the center of the core:
$$
\rho_{n0}=\frac{M_n}{4\pi}\left\{\int_0^{\infty}
exp \left(-\frac{\psi(r) m_x}{T^{\star}}\right)r^2 \, dr \right\}^{-1}
$$
 Equations (13) take into account that the crossection of
CDM particles with baryons is small enough. It means that these particles
oscillate many times inside the core between collisions with baryons.
Using (15) we obtain CDM term in equations (13)
$$
\frac{dP_n}{dr}= \frac{T^{\star}}{\rho_n}\,\frac{d\rho_n}{dr}
$$

Equations (13) have two different solutions describing two different
stationary states of the baryonic core containing nonbaryonic component.
The first one is usual, determined by baryons in the whole range of $r$:
$0\leq r \leq r_b$. It has value of baryons density in the center $\rho_b
|_{r=0} = \rho_{b0}$ and distribution
$\rho_b(r)$ gradually falling down to the edge
of the body $\rho\to 1$, $r\to r_b$. For the planet of Jupiter type
$\rho_{b0}\approx 30$ g/cm$^3$, but the temperature inside Jupiter is
essentially less than in the center of BC.
In our case  the mass of baryonic body $M_b
\sim 0.05M_x$ could be higher, than Jupiter mass, but taking into
account temperature,
one can expect, that central baryonic density by the order of magnitude is
\begin{equation}
\rho_{b0}\sim 10 \, \frac{g}{cm^3}
\end{equation}
If the density of nonbaryonic component $\rho_n(r)$ is everywhere less, than
$\rho_b(r)$
\begin{equation}
\rho_n(r)<<\rho_b(r)
\end{equation}
the existence of CDM particles, trapped inside BC will lead to the small
corrections to the usual baryonic stationary state.

But according to the Boltzman distribution (14)
\begin{equation}
\rho_n=\rho_{n0}\,
exp \left\{-\left(\frac{r}{r_{n0}}\right)^2\right\}
\end{equation}
Here $r_{n0}$ gives the characteristic scale of the heated region
\begin{equation}
r_{n0}=
\sqrt{\frac{3 T^{\star}}{2\pi G \rho_{b0} m_x}}
\end{equation}
In (18),(19) we supposed a constant baryonic density $\rho_b=\rho_{b0}$
. Density of CDM particles in the centrum of the core $\rho_{n0}$ is
connected with full mass $M_{xb}=M_n$ by relation:
\begin{equation}
\rho_{n0}=\frac{M_n}{\pi^{3/2} r_{n0}^3}
\end{equation}
Condition (17) is fulfilled only if
$$
\frac{M_n}{\pi^{3/2} r_{n0}^3}< \rho_{b0}
$$
Taking into account (19) we see, that for low temperatures $T^{\star}$
or for the high mass of CDM particles if:
\begin{equation}
M_n\,\rho_{b0}^{1/2}\left(\frac{2\,G\,m_x}{3\,T^{\star}}\right)^{3/2}>1
\end{equation}
condition (17) is not fulfilled. In this case $\rho_{n0}>\rho_{b0}$ and CDM
particles could significantly affect the density distribution
being trapped by their own gravitational field
in the central part of BC. General stationary solution of equation
(13) is shown in the Fig.1. One can see from the figure that two
different stationary modes for potential $\psi$ exist. One  is the
mode determined mostly by baryons, this mode we have discussed
earlier. Another is the CDM particles selftrapped mode. In the vicinity
of the centrum $r\to 0$ the last mode is asymptotically described
by:
\begin{equation}
\psi=\frac{2T^{\star}}{m_x}\,ln\left(\frac{r}{R_c}\right) + ...
\end{equation}
$$
\rho=\frac{T^{\star}}{2\pi m_x G r^2}+...
$$
Here $R_c$ determines the dimension of the region of selftrapped
CDM bulk
\begin{equation}
R_c=\frac{G\,M_cm_x}{2\pi T^{\star}}
\end{equation}
$M_c$ is the full mass of selftrapped neutralino. The main  term
in (22) coincides with well-known Boltzman distribution of noninteracting
gravitationally selftrapped matter [18].

The existence of a second stationary state means that if in the process of
the slow cooling of the core conditions (21) could be reached
then the transformation
to the second stationary state will begin. This reconstruction will lead
to the loss of the hydrodynamic stability of the system
resulting in the release of a significant amount of energy $E\sim 10^{43}
- 10^{44}$ erg (see (35),(36)).
This energy will be emitted into a wide spectrum of electromagnrtic waves
, including gamma rays, X-rays, optic emission and radio waves.

\section{Explosion of baryonic core in neutralino star}.

The particles of which CDM is composed are not known now, though
there exist some hypothetical candidates: neutralino, heavy
neutrino, axions, strings. Neutralino and heavy neutrino like
the Majorana particles can annihilate in mutual collisions.
Noncompact objects which consist of such particles are
partially disappearing
 because of annihilation processes. These specific objects
we call "neutralino stars" (NeS) [12] and for nonbaryonic particles
from which they consist of we will use
as a general name "neutralino" [13].

Annihilation of neutralino leads to effective radiation of gamma
photons and
heating of baryonic core. Under definite conditions the core can
explode. It should be emphasized, that the
energy released in all these processes was input into CDM during
the period of its freezing out. The possibility to detect CDM
particles through their annihilation was discussed in a number of
papers (see [19,20]). In these papers always was supposed a smooth
distribution of CDM in the Galaxy.
The existence of NeS leads to a new situation
, connected with compression of CDM in NeS, which
results in strong amplification
of neutralino annihilation processes and gives significant constrain
on possible type of neutralino particles [13,17,21].
From the other side it opens the
possibility of existence of quite observable fluxes of gamma radiation
by NeS. In particular, in [17]
a NeS model of diffusive gamma emission
and recently discovered nonidentified point like gamma sources [22] is
developed.  Here we will discuss a  model of gamma bursts, which is connected
with the possibility of blowing up of the baryonic core of NeS.

As was shown in section 2, the amount of the trapped in BC neutralino
particles is high enough $M_{xb}\sim (10^{-7}-10^{-6}) M_x$. Because
of this reason the energy released in neutralino annihilation could
effectively heat the baryonic core. On the other hand, the neutralino
density  $\rho_n$
could be small in comparison with the density of baryons $\rho_b$.

The heating term $Q_n$ in  equation (13) could be written in a form
\begin{equation}
Q_n=2\,\frac{\rho_n c^2}{\tau_n}
\end{equation}
where $\rho_n(r)$- is the neutralino density (18)-(20), and $\tau_n$ is
the lifetime of neutralino in baryonic core. Taking into account that
neutralino annihilation is a rare event in comparison with neutralino
collisions with baryons, we obtain the following expression for $\tau_n$
\begin{equation}
\tau_n= \frac{m_x}{\rho_{n0}<\sigma v>}
\end{equation}
Here $<\sigma v>$ is a characteristic crossection of neutralino annihilation
multiplied on typical velocity. Note that $\tau_n$ does not depend on $r$.
So, the heating efficiency $Q_n$is directly proportional to $\rho_n(r)$.

The distribution of $\rho_n(r)$ is given by equation (18). It follows from (24),
that if conditions (21) are fulfilled the heating is going mostly in the
central region of the core.

Let us estimate now the main parameters of the core. From (19) it
follows that:
\begin{equation}
r_{n0}=2.4\times 10^9 \left(T^{\star}/10^6 K^o\right)^{1/2}
\left(10 \, Gev /m_x \right)^{1/2} \left(\rho_{b0}/10 \,g\,cm^{-3}
\right)^{-1/2} \, cm
\end{equation}
One can see that the dimension of heated region could be not too small
in comparison with the radius of baryonic core:
\begin{equation}
R_b=\left(\frac{3 \, M_b}{4\pi \, \rho_b}\right)^{1/3}\approx
2\times 10^{10} \, cm
\end{equation}
Here and below we take:
$$
M_b=0.05M_x=5\times 10^{31}\,g, \qquad M_x=0.5M_{\odot}=10^{33}\,g,
\qquad \rho_b=<\rho_b>=1\,\frac{g}{cm^3}
$$
Density of neutralinos in the center of BC, as follows from (20) is:
\begin{equation}
\rho_{n0}= 1.3\times 10^{-4}\left(\frac{M_n}{10^{25} g}\right)
\left(\frac{10^6 K^o}{T^{\star}}\right)^{3/2}\left(\frac{m_x}{10\, Gev}
\right)^{3/2}\left(\frac{\rho_{b0}}{10\, g/cm^3}\right)^{3/2}\,\frac{g}{cm^3}
\end{equation}
and their number density:
\begin{equation}
n_0= 7.6\times 10^{18}\left(\frac{M_n}{10^{25} g}\right)
\left(\frac{10^6 K^o}{T^{\star}}\right)^{3/2}\left(\frac{m_x}{10\, Gev}
\right)^{1/2}\left(\frac{\rho_{b0}}{10\, g/cm^3}\right)^{3/2}\, cm^{-3}
\end{equation}
We see that conditions (17), is usually fulfilled.

Characteristic annihilation time for neutralinos in the core could be
of the same order as the lifetime of the Universe:
\begin{equation}
\tau_n=3.2\times10^{15}\left(\frac{3\times 10^{-27}}{<\sigma v>_0}
\right)\left(\frac{10^6K^o}{T^{\star}}\right)^{-1/2}\left(
\frac{10^{25}g}{M_n}\right) \left(\frac{\rho_{b0}}{10\,g/cm^3}\right)^{-3/2}
\left(\frac{10\,Gev}{m_x}\right)^{-1/2}\, s
\end{equation}
Here we took into account that according to [13], the dominant annihilation
process of neutralinos is $p$-wave  with crossection:
\begin{equation}
<\sigma v>=<\sigma v>_0 \frac{3\,T^{\star}}{2\,m_x c^2}\,,
\qquad <\sigma v>_0=\left(10^{-26} - 10^{-27}\right)\left(\frac
{m_x}{10 \hbox{Gev}}\right)^2\, \frac{cm^3}{s}
\end{equation}
The annihilation time of neutralino should be greater than life-time of
the Universe $t_0$. It depends very strongly (according to (30)) on the
parameters of the NeS. If it is not so, the mass of neutralino $M_n$ will
decrease due to annihilation until $\tau_n \approx t_0$. It means that
energy  release  due to annihilation which heats
central region of the core in present time is (24):
\begin{equation}
Q_{n0}= 6.8\times 10^{-2}
\frac{erg}{cm^3 s}
\end{equation}
Consequently the full energy dissipated by neutralino in BC is:
\begin{equation}
Q=Q_{n0}r_{n0}^3\pi^{3/2}=1.4\times 10^{29}\, \frac{erg}{s}
\end{equation}
This energy being transported to the surface of the core is emitted
due to usual blackbody radiation.
\begin{equation}
Q=Q_s=\sigma T_s^4 S =1.4\times10^{29}\left(\frac{T_s}{800^oK}
\right)^4\left(\frac{R_b}{2\times 10^{10}\,cm}\right)^2 \,\frac{erg}{s}
\end{equation}
Relation (33) determines the surface temperature $T_s$ as a function
of parameters. For the stationary state formulas (25)-(33) describe
some average situation, the exact distribution of
temperature would be established in accordance with equation (13).
To fulfill the
stationary conditions  a convective heat transport near the boundary region has
to be selforganized.

Let us estimate the hydrodynamic stability  (13) comparing the full
pressure of baryonic gas $P$ with the gravitational energy $E$. Taking
$\rho_{b}$ constant we have:
\begin{equation}
E=\frac{3\,G\,M_b^2}{10\,R_b}
\end{equation}
were $M_b$ is the full mass of the baryonic core. Total particle
energy
$$
P=N \bar{T}
$$
where $\bar{T}$ is the average temperature and $N$ - number of baryonic
particles. Here we took into account that the temperature is high
enough to keep the object in gaseous state. Taking $\bar{T}\approx
0.6 T^{\star}=6\times  10^{5} K^o$, one can obtain
\begin{equation}
P\approx E=2.5\times 10^{45}\left(\frac{M_b}{5\times 10^{32}\,g}
\right)^2 \, erg
\end{equation}
We see that the particle energy is quite close to the gravitational
one, it means that the main integral relation for equation (13) is
well fulfilled.

So the structure of baryonic core of NeS because of neutralino
heating is more alike to the star than to the planet.
In the stationary conditions it has the temperature in the
central part of the core an order of $10^6 K^o$. But the
surface temperature is not very high.

We emphasize that the time needed to reach stationary
temperature distribution in BC is compatible with the age of
Universe. Because of this reason nonstationary processes could be
significant. As the heating is going in the central part  the
temperature in this part
$T^{\star}$ at some conditions could be much higher than
 the stationary one. Nonstationarity of $T^{\star}$
depends on the thermal conductivity and grows especially strongly
with the full mass of neutralinos $M_n$ trapped in the core.
The strong growth of
$T^{\star}$
follows from the relation (24) for neutralino dissipation rate and
if $M_n\geq 10^{26}$ g it could increase on two orders of magnitude
or even higher.
This process can lead to the effective heating of the small central
part of the core . The temperature in the center of BC could then increase
in nonstationary process up to
\begin{equation}
T_m\approx 10^7 \, K^o.
\end{equation}

So we see that the baryonic core of NeS because of the heating by
neutralinos is quite peculiar astrophysical object which could be
constructed either like a planet or like a star depending on the
efficiency of heating. The core can lose its hydrodynamical stability
and explode, releasing the energy of the order of (35)
$$
E\sim (0.1-0.01)\,P\sim 2\times\left(10^{43} - 10^{44}\right) \, erg
$$
This energy is enough for creation of gamma burst
($E_{\gamma}\sim 10^{41}-10^{42}$ erg), just what is needed for GDMH
model of gamma bursts. We will consider now three main destabilization
factors which can lead to explosion of BC.

1. Overheating of the central part of BC by neutralino.

A very strong heating of BC is connected with the high value of
bulk of trapped neutralinos $M_n\geq 10^{26}$ g. The heating would
be more concentrated in the centrum of BC for the high values of
neutralino mass $m_x$. As the surface temperature of BC is of the
order $500^o-1000^o$ (what follows from the conservation of thermal
flux) it is only weakly ionized what leads to the strong depression
of thermal conductivity. Convective transport would be the main
process of the heat transport in this region. Overheating in the central
region could lead to the loss of hydrodynamical stability and explosion of BC.

2. Thermonuclear heating.

Neutralino annihilation inside the star leads to generation of
highly energetic protons and gamma photons. These energetic
particles produce ions of $D_2^+$ and some other light
elements by interaction with the bulk
protons of the core. Because of this process the number of
$D_2$ and other particles is constantly growing in time
in the central part of BC. Following (24), one can estimate the number
density of deuterium produced for the lifetime of the Universe:
\begin{equation}
N_{D_2}=
\gamma_{D_2}\frac{Q_{n0}}{m_xc^2}t_0
\end{equation}
$$
N_{D_2}=2\times 10^{21}\gamma_{D2}\left(\frac{M_n}{10^{25}g}
\right)^2\left(\frac{10^6\,K^o}{T^{\star}}\right)^{1/2}
\left(\frac{m_x}{10\,Gev}\right)^{5/2}\left(\frac{\rho_0}{100\,g\,cm^{-3}}
\right)^{3/2}\left(\frac{<\sigma v>_0}{3\times 10^{-27}}\right)\,cm^{-3}
$$
$\gamma_{D_2}$ is the transformation coefficient for Gev-energy
protons into $D_2$ ions due to collisions with the bulk protons.
The number density (37) is big enough for initiation
$D_2\to He_{3}\to He_{4}$ thermonuclear reaction.
It means that if the
temperature could reach high values (36) the effective nuclear
process will begin.
This process determine the heating term $Q_{th}$ in equation (13)
Thermonuclear process leads to a very effective heating $Q_{th}$ (13)
of the central part of BC and could become explosive
, depending on initial state. It is necessary to mention that in (37) the
total mass of trapped neutralino should be take into account, because
even if according to (30) netralino annihilate, the process of production
of $D_2$ will take place.

3. Selftrapping of neutralino.

This process has been already considered in the previous section.
As was mentioned above the dimension $R_c$ could be small, it is diminishing
with the growth of the mass of neutralino particles $m_x$. On the other
hand to reach the conditions (21), which determine the possibility of
formation of selftrapped neutralino bulk
(SNB) easier for low temperatures.  So, it could be that
SNB is formed during nonstationary process, when temperature in BC
reach its minimum.

The fundamental feature of SNB is that the gravitational selftcompression
is growing up with the growth of temperature $T^{\star}$: its dimension
became smaller (23) and neutralino density becomes higher (22).
Such process leads to an explosive heating of the central region
of BC. As follows from (22),(23),(24),(31) the energy,
dissipated in SNB in 1 cm $^3$
per second is:
\begin{equation}
Q_{n1}=3.2\times 10^{32} \left(\frac{T^{\star}}{10^6K^o}\right)^7
\left(\frac{10\,Gev}{m_x}\right)^8 \left(\frac{10^{25}\,g}{M_n}
\right)^4 \left(\frac{<\sigma v>_0}{3\times 10^{-27} cm^3/s}
\right)
\left(\frac{R_c}{r}\right)^4 \, \, \frac{erg}{cm^3 s}
\end{equation}
And the full dissipated power is:
\begin{equation}
Q=2.5\times 10^{44} \left(\frac{R_c}{r_{min}}-1\right)
\left(\frac{T^{\star}}{10^6K^o}\right)^4
\left(\frac{10\,Gev}{m_x}\right)^5 \left(\frac{10^{25}\,g}{M_n}
\right) \left(\frac{<\sigma v>_0}{3\times 10^{-27} cm^3/s}
\right)\, \, \frac{erg}{s}
\end{equation}
Were $r_{min}$ is a minimal dimensions for the bulk.
We see, that
the energy  dissipation in SNB is explosive, lasting only a few seconds.

It is evident, that to reach in reality such a stationary state is impossible
, but nonstationary transition to this state is
also a fast process, which can explode the BC. Let us determine
the characteristic time scale $\tau_n$ of this process.
Neutralinos crossection for elastic collisions is usually lower
than for annihilation. So, they can establish Boltzman distribution
only through collisions with baryons. Characteristic crossection
for neutralino-proton collision according to [23] is:
$$
<\sigma_{np} v>\approx 10^{-29}\left(\frac{\rho_b}{1\,g/cm^3}
\right)
\, \, \frac{cm^3}{s}
$$
Supposing the density of protons (16) $\rho_b\approx 10\,g/cm^3$  we
obtain
$$
\tau_n\sim 10^7 s
$$
This time is much smaller than the characteristic time of establishing
of thermal equilibrium by the heat transport. So we conclude, that it
is possible to have an overheating and blowing up of BC due to
selftrapping, heating and fast compression of neutralino bulk in the
centrum.

\section{Conclusions}.

1.We have shown here that MACHO's in the form of
noncompact CDM halo objects could serve as the
source of gamma bursts. This model could be proved if in experiments
the significant local asymmetry of gamma bursts distribution, connected
with the existence  of a Giant Halo of the Andromeda (M31) would be
established by observations. To do that a device with the sensitivity
an order of magnitude higher than BATSE is needed. These experiments
were already proposed earlier in connection with GDMH model [5].

2. The explosion of barionic core of NO proposed here as a source
of gamma bursts will lead also to the emission of X-rays, optic
infrared and radio waves. So a wide spectrum of electromagnetic
emission follows the gamma burst.

3.Let us discuss now the interesting new opportunities for
observations which follow from our model.
The intensive infrared emission from BC of neutralino star (NeS) is predicted
in present work according to formulae (34).
The wave length of this emission $\lambda\sim (3 - 8)\mu$, its full
power  $Q\sim 10^{28} - 10^{29}$ erg/s. Emission should have blackbody
spectrum with effective temperatures $T_s\sim(500 - 1000)K^o$. The intensity
of this emission at the Earth from the nearest sources should be of the order
$S\sim (10^{-10} - 10^{-11})$ erg/s cm$^2$.

We emphasize, that the theory
predicts gamma emission from the same NeS also. In [17] are distinguished
nonidentified point-like gamma sources from EGRET data [22]. It is shown
that these sources could be considered as the emission generated by NeS.
A more detailed analysis [15], demonstrated the existence of two types of
the NeS gamma sources. Galactic type has the emission intensity an order
of magnitude higher, than halo type, considered in [17], what is in full
agreement with the intensity of point-like gamma sources observed by
EGRET [22]. In [15] demonstrated also an especially good agreement of a new
EGRET observations with the predictions of a space distribution of sources
made in [17], basing on NeS model of gamma sources.  So of a great interest is
to observe the infrared emission from the same sources. If this observations
would be successful it would mean that we see the neutralino stars: neutralino
halo of such a star radiates in gamma and its baryonic core in infrared rays.

We conclude, that infrared observations of nonidentified point-like gamma
sources [17] are of fundamental significance as they could lead to the
discovery of NeS.

\vspace{0.5cm}

The authors are grateful to V.L.Ginzburg for permanent interest and
fruitful discussion, and to
I.Axford, V.A.Dogiel,
A.Lukyanov and V.Sirota for valuable remarks.

We are grateful for hospitality to  the Max Planck Institute of Aeronomy,
where a part of this work was fulfilled and to
 Russian Fund of Fundamental Research \\
(Grant  96-02-18217), which support our work.

\vspace{0.5cm}

{\bf References}

\vspace{0.5cm}

1. G.J.Fishman et al Astrophys.J. 461 (1996)  84.

2. A.V.Gurevich,V.S.Beskin,K.P.Zybin,M.O.Ptitsyn Proc.of IV-th
Int.Conf.on Plasma

Phys. and Cont.Nuclear Fussion, Toki, Japan,
17-20 November 1992 (ESA SP - 351).

3. A.V.Gurevich,V.S.Beskin,K.P.Zybin,M.O.Ptitsyn
Sov.Phys. JETP  103, (1993), 1873

4. A.V.Gurevich,V.S.Beskin,K.P.Zybin,M.O.Ptitsyn
Phys.Lett.A 175 (1993), 397

5. A.V.Gurevich,G.F.Zharkov,K.P.Zybin,M.O.Ptitsyn
Phys.Lett.A 181 (1994) 289

6. K.Bulik, D.Q.Lamb AIP Conf.Proc.Conf. on High Velocity Neutron Stars,
ed. by

R.Rothshild, in press

7. P.Podsiadlowski, M.J.Rees, M.Ruderman Mon.Not.Roy.Astr.Soc, 273 (1995), 755

8. F.X.Timmes, S.E.Woosley, T.A.Weaver Astroph.J.Sup. 98 (1995), 617

9. D.H.Hartmann, R.Narayan Astrophys.J. 464 (1996), 226

10. C.A.Alcock, et al. Nature 365 (1993), 621

11. M.R.Pratt, et al (The MACHO collaboration) preprint astro-ph/9606134.

12. A.V.Gurevich and K.P.Zybin, Phys.Lett.A 208 (1995) 267.

13. A.V.Gurevich,K.P.Zybin,V.A.Sirota, Phys.Lett.A 214 (1996) 232

14. A.V.Gurevich,K.P.Zybin,V.A.Sirota, II Int.Sakharov Conference, Moscow
(1996)

15. A.V.Gurevich,K.P.Zybin,V.A.Sirota, Sov.Phys.Uspekhi (1997) in press

16. A.V.Gurevich,K.P.Zybin, Sov.Phys.Uspekhi 38 (1995) 687

17. A.V.Gurevich and K.P.Zybin, Phys.Lett.A  225, 217 (1997)

18. P.J.E.Peebles The Large Scale Structure of the Universe (Prineceton
University Press,

1980)

19. L.Roskowski, Phys.Lett.B 278 (1992) 147.

20. A.Botino et al Astroparticle Phys. 2 (1994) 77

21. V.Berezinsky, A.Botino, G.Mignola preprint CERN-TH/96-283 October 1996.

22. D.J. Thompson et al Astrophy.J.Suppl. 101 (1995) 259.

23. A.Botino, et al. Astroparticle Phys. 2 (1994), 77

\newpage
\rule{0pt}{2cm}

\begin{center}
Fig.1
\end{center}

Two solutions of equation (13). First one correspond to neutralino
selftrapping regime (curve 1) at the center of barionic core, and
the second one is regular solution defined basically by barions
density (curve 2).

\end{document}